# The Surface of Electrolyte Solutions is Stratified

Yair Litman[1,2]*, Kuo-Yang Chiang[1], Takakazu Seki[1],

Yuki Nagata[1], Mischa Bonn[1]*

Max Planck Institute for Polymer Research,

Ackermannweg 10, 55128 Mainz, Germany

Yusuf Hamied Department of Chemistry, University of Cambridge,

Lensfield Road, Cambridge, CB21EW, United Kingdom

*To whom correspondence should be addressed:
E-mail: litmany@mpip-mainz.mpg.de, bonn@mpip-mainz.mpg.de

The distribution of ions at the air/water interface plays a decisive role in many natural processes. It is generally understood that polarizable ions with low charge density are surface-active, implying they sit on top of the water surface. Here, we revise this established hypothesis by combining surface-specific heterodyne-detected vibrational sum-frequency generation with neural network-assisted *ab initio* molecular dynamics simulations. Our results directly demonstrate that ions in typical electrolyte solutions are, in fact, located in a subsurface region leading to a stratification of such interfaces into two distinctive water layers. The outermost surface is ion-depleted, and the sub-surface layer is ion-enriched. As a result, an effective liquid/liquid interface buried a few Å inside the solution emerges, creating a second water/electrolyte interface, in addition to the outermost air/water interface.

Around 70% of the earth's surface is covered by ocean water. The evaporation and heterogeneous aerosol formation of such electrolyte-rich solutions play an important role in atmospheric chemistry and climate science[1–3]. The physicochemical processes occurring on the surface of electrolyte solutions are ultimately controlled by the local electric fields and molecular structure of such solutions at the air/liquid interfaces. Thus, a microscopic knowledge of the ion distributions and molecular orientation at the arguable simplest air/solution interface is of paramount importance for the development of environmental models [2,4], and also serves as a starting point for understanding more complex interfaces of liquid solutions in contact with electrodes, membranes, or minerals [5–7].

The accepted molecular picture of the surface of aqueous electrolyte solutions is based on results obtained with surface-specific spectroscopic techniques and classical molecular dynamics (MD) simulations[8–11]. Briefly, polarizable heavy ions, such as $Br^-$ and $I^-$, accumulate at the interfacial region, while less polarizable ions, such as $F^-$, $Na^+$, are depleted from the interface and remain buried in the solution. This differential distribution of anions and cations is presumed to create the so-called electric double layer (EDL), an ionic double-layer structure, or simply double layer[8,9,12–14]. The application of the EDL model to air/liquid interfaces has played a fundamental role in describing surface chemistry. The EDL model has replaced earlier views in which simple inorganic salts were thought to be depleted from the interface[11,15].

It is challenging to obtain molecular-level insights into the local ion distribution and water structure at the surface of electrolyte solutions. Vibrational sum-frequency generation (VSFG) is a surface-specific technique that directly probes the response from molecular vibrations interfaces. Atomic ions support no vibrations, so their behavior can only be inferred indirectly from the VSFG response of the water, typically invoking the EDL picture. VSFG signal intensity variations in the 3000-3600 $cm^{-1}$ region upon adding salt have been attributed to excess accumulation of cations or anions at the interface[8,13,14,16–18]. Yet, discrepancies have recently become evident, and a unifying picture is absent in the literature. For example, while most polarizable force field-based calculations predict a strong enhancement of the I⁻ concentration at the interface[20], in agreement with measured surface excess free energy[21], measurements reported by Raymond and co-workers[22] concluded a lack of significant surface enhancement of polarizable anions through a careful design of isotopic dilution experiments, in agreement with more recent *ab initio* simulations [23].

Progress in this area is restricted by several challenges: (i) low SFG signal levels from these types of interfaces; (ii) the necessity for determining the signal phase (by heterodyne detection) for recording unambiguous data; and (iii) the insufficiency of experimental data alone to disentangle spectral components in aqueous solutions unambiguously, as the water VSFG response is often broad and featureless[19].

In this work, we overcome these challenges by combining high-level heterodyne-detected VSFG (HD-VSFG) data with neural network (NN)-aided *ab initio* MD (NN-AIMD) simulations and, to obtain a unifying picture, study the structure of the interface of several electrolyte solutions, namely HCl, NaOH, CsF, NaCl, NaBr, and NaI. HD-VSFG gives direct access to $\text{Im}(\chi^{(2)})$ and allows disentangling the non-resonant background in the HD-VSFG spectra revealing subtle but essential details in the actual resonant spectra. By systematically changing the salt concentration and comparing HD-VSFG spectra with simulated spectra directly, we show that the EDL model can partially explain the HD-VSFG spectra of the sodium halide salts but fails for other cases, the only exception being HCl. Combined experimental data and simulations with *ab initio* quality reveal a liquid/liquid interface, a few angstroms below the air/electrolyte solution interface. Such a second interface represents a refined, more accurate description of the interfacial structure of electrolyte solutions.

## Results and Discussion

Figure 1 shows the $\text{Im}(\chi^{(2)})$ spectra for water and different electrolytes solutions (panels A-D). First, we focus on the spectrum at the water-air interface. The spectrum of pure water shows a sharp positive peak centered at 3700 cm$^{-1}$ and negative amplitude signal around 3200-350 cm$^{-1}$. The 3700 cm$^{-1}$ peak arises from the free O-H group of the topmost interfacial water, while the 3200-3550 cm$^{-1}$ band arises from the hydrogen-bonded (H-bonded) O-H group of the interfacial water. The positive (negative) sign of the O-H stretch mode indicates that the O-H group points *up* to the air (*down* to the bulk).

When compared to the data of pure water, the spectrum of the HCl solution shows a decrease of the free O—H stretch band (less positive) and a decrease of the H-bonded stretch band (more negative). The decrease of the free O—H peak is an unambiguous indication that protons reside on the topmost layer by capping the free O—H groups with the hydronium ions. The surface propensity of protons has been confirmed experimentally [24,25] and theoretically [26] in the past. In contrast to the HCl solution case, the $\text{Im}(\chi^{(2)})$ spectra for the NaCl, NaOH, and CsF solutions show that these ions modify the H-bonded band by enhancing and/or reducing their amplitudes, but the free O—H band remains largely unaffected. The unchanged free O-H group manifests that the free O-H groups of the interfacial water molecules are not capped by these ions.

Panels E-G in Figure 1 show the concentration dependence of the 3200 cm$^{-1}$, 3450 cm$^{-1}$, and 3700 cm$^{-1}$ amplitudes (the free O—H peak, and the high- and low-frequency sides of the H-bonded O—H stretch bands, respectively). Cl$^-$, OH$^-$, and F$^-$ ions do not affect the free O-H peak amplitude significantly within the considered concentration range, unlike the hydronium ions. In contrast, the H-bonded O-H band amplitudes change significantly for all the ions, although with different magnitudes and signs. These results question the use of one unique model to describe the diverse behavior of ions at the interface.

Note that our $\text{Im}(\chi^{(2)})$ spectra of the NaCl and NaOH solutions, as well as the pure water, differ from the reported data particularly in the <3200 cm$^{-1}$ region[16,17,27,28]. Phase inaccuracies in the early measurement created an artificial positive peak below 3200 cm$^{-1}$. Thanks to the recent procedure of accurate phase determination, we could obtain the phase accurately in our HD-VSFG spectra. Indeed, our air/water interface data agrees with the recent reports[27,29].

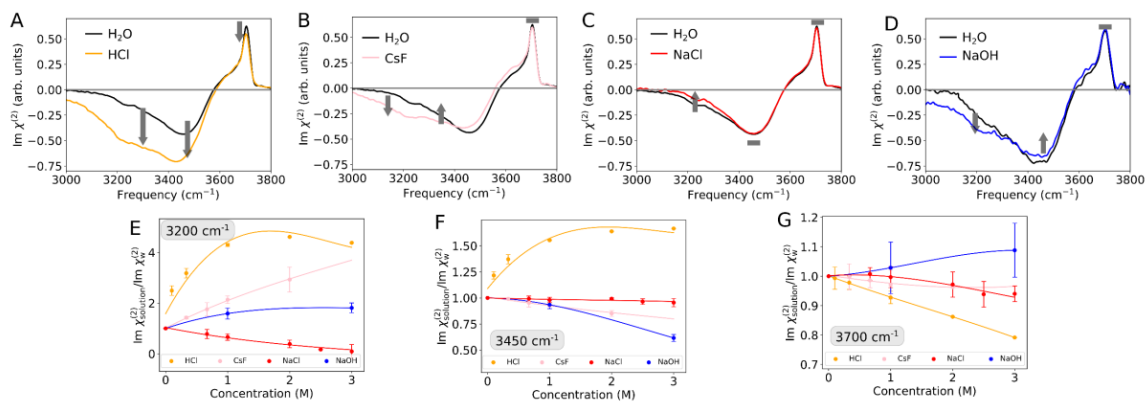

**Figure 1: Experimental HD-VSFG spectra at air/electrolyte solutions interface.** HD-VSFG spectra for **(A)** HCl 1M solution (orange), **(B)** CsF 1.5M solution (pink), **(C)** NaCl 1.5M solution (red), and **(D)** NaOH 1.0M solution (blue). The free O—H and H-bond bands are altered with respect to pure water reference for the HCl case (A), while for the CsF and NaCl cases, only the H-bonded bands are modified. In all panels, the water VSFG spectra of water (black) are shown for comparison. The direction of the spectral changes is highlighted with arrows, and gray rectangles depict negligible spectral changes. Concentration dependence of the HD-VSFG amplitudes, expressed as the ratio of electrolyte solution and water signals (Im ($\chi^{(2)}_{\text{solution}}/\chi^{(2)}_{\text{w}}$)), at 3200, 3450, and 3700 cm$^{-1}$ are shown in panels **(E)**, **(F)** and **(G)**, respectively. Lines represent a polynomial fit used to guide the eye without physical meaning.

To explore why the free O-H peak is unchanged, although the H-bonded band is changed upon the addition of salt, we carried out theoretical calculations of VSFG spectra. Among NaCl, CsF, and NaOH, we decided to investigate the NaOH spectrum in more detail by simulating its HD-VSFG spectra, because NaOH serves as a representative of the electrolytes that do not perturb the free O—H band, and it shows the most drastic change of the HD-VSFG features. The calculation of VSFG spectra for water is at least 10 times more expensive than the bulk spectroscopies such as Raman or IR[30], since the signal comes from only very few water layers, and the signal arising from bulk should correctly be averaged to zero. Moreover, the description of bond breaking and formation events, as well as the need for accurate spectra, demand to go beyond classical force fields and perform *ab initio* calculations[19,23]. The latter demands a considerably greater computational effort, which explains why *ab initio* calculations of SFG spectra are scarce in the literature. To overcome this daunting computational cost, we carried out neural network (NN)-aided *ab initio* MD (NN-AIMD) simulations[31,32], which allowed us to perform the required simulations at a fraction of the computational cost, while preserving the underlying quantum mechanical accuracy.

Figures 2a and 2b shows the experimental and theoretical HD-VSFG data, respectively. The comparison of the experiment and simulation shows that our simulations capture the two main spectral changes correctly. For both set of spectra, the addition of NaOH ions triggers a decrease in the H-bonded negative signal between 3300 $cm^{-1}$ and 3600 $cm^{-1}$ – it becomes less negative – and the emergence of a broad continuum negative signal in the 2400-3200 $cm^{-1}$ region. The free O-H remains unperturbed. The simulations capture the main trends observed experimentally and reproduce the described spectral changes with increasing salt concentration. It also lacks the shoulder at 3600 cm$^{-1}$ arising from the asymmetric stretching mode in water molecules that donate two hydrogen bonds and accept one[33], due to the well-known limitations of the surface-specific velocity-velocity correlation function (ssVVCF) approximation[30]. Unlike previous simulations based on classical force-fields[34], our NN-AIMD results correctly predict no SFG signal for pure water below 3000 cm$^{-1}$. Further validation tests are given in the supporting information (SI).

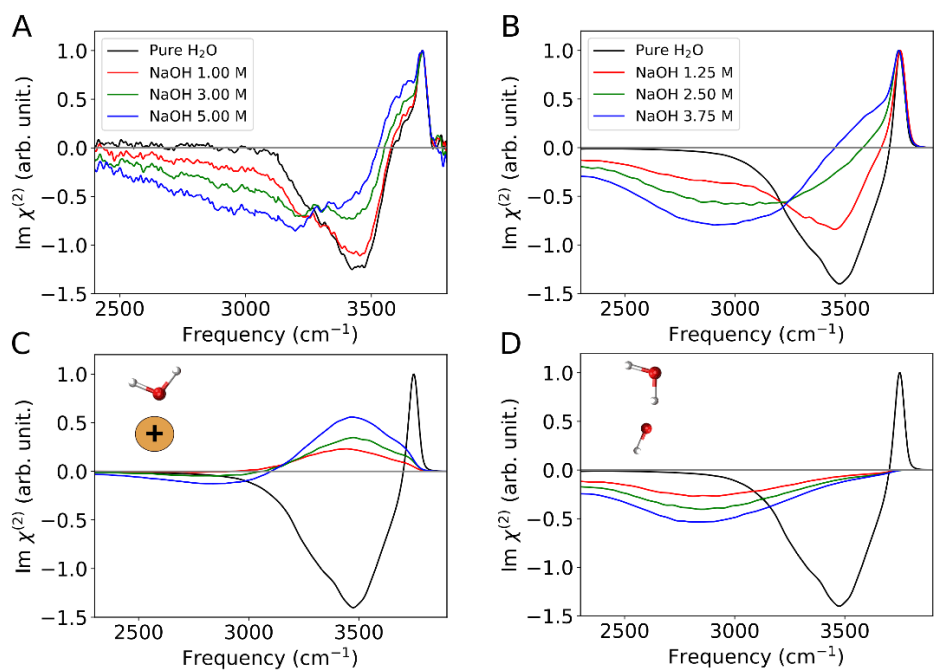

**Figure 2**: **Experimental and theoretical VSFG spectra of NaOH aqueous solutions at room temperature (A)** Experimental imaginary part of $\chi^{(2)}$ spectra obtained for water (black), NaOH 1.00 M (red), NaOH 3.00 M (green), and NaOH 5.00 M (blue) aqueous solutions. **(B)** Theoretical prediction based on the ssVVCF methodology[30]. **(C)** Contribution to the theoretical spectra by the water molecules in the vicinity of Na$^+$ ions **(D)** Contribution to the theoretical spectra by the OH bonds coordinating OH$^-$ anions. The pure water spectrum is shown in all panels in black as a reference. The theoretical spectra have been corrected by estimation of the frequency-dependence of the transition dipole and polarizability moments[35,36]. A reference zero line has been added in gray.

Having validated our theoretical framework, we proceed with an analysis of the simulated spectra. Figures 2c and 2d present the $\text{Im}(\chi^{(2)})$ spectra contributed by selected water molecules in the proximity of Na$^+$ and OH$^-$, respectively. Water molecules coordinated to Na$^+$ ions contribute to a positive signal between 3200 and 3800 cm$^{-1}$, while the O—H bond of water molecules coordinated to OH$^-$ anions are responsible for the negative < 3500 cm$^{-1}$ continuum. The latter assignment is consistent with previous studies of NaOH in bulk[37–39] and water clusters[40,41] where the <3500 cm$^{-1}$ continuum band is also attributed to the solvation of OH$^-$ species. Note that the continuum that extends below 2500 cm$^{-1}$ does not arise from the bulk $\chi^{(3)}$ contributions because the bulk $\chi^{(3)}$ contributions have only a small contribution below 3000 cm$^{-1}$, unlike the spectra shown in Figure 2d.[42,43]

This spectral decomposition strongly indicates that the HD-VSFG spectra at the electrolyte air/solution interface ($\chi^{(2)}_{\text{solution}}$) can be described as the sum of the VSFG spectrum of (unperturbed) water at the air/water interface ($\chi^{(2)}_w$), of water oriented by the cation ($\chi^{(2)}_c$), and that of water oriented by the anion ($\chi^{(2)}_a$);

$$\chi^{(2)}_{\text{solution}} = \chi^{(2)}_w + c\chi^{(2)}_{\text{electrolyte}} = \chi^{(2)}_w + c(\chi^{(2)}_c + \chi^{(2)}_a), \qquad (1)$$

where $c$ is a parameter proportional to the salt concentration, and we have defined the electrolyte contribution, $\chi^{(2)}_{\text{electrolyte}} = \chi^{(2)}_c + \chi^{(2)}_a$, for later convenience. Note that $\chi^{(2)}_{\text{electrolyte}}$ also includes the effect due to water displacement.

This interpretation differs from the conventional EDL in which cationic and anionic contributions are entangled; within the EDL picture, it is assumed that the local electric field created by the charge separation at the interface induces a net average orientation of the O—H transition dipole moments of the water molecules in the topmost layers. However, such a scenario cannot account for the simultaneous decrease and increase in the signal around 3000 and 3500 cm$^{-1}$, respectively. Two spectral changes with contrary signs would require alignment with opposite orientations, representing a distinct limitation of the EDL model.

To further explore the ion distribution near the interfaces, we computed the density profiles for the different species, which are displayed in Figure 3a. Both Na$^+$ and OH$^-$ ions are repelled from the interface, giving rise to an ion-depleted topmost layer – pure in water. In Figure 3b, we present the orientation profile for water, OH$^-$, and water molecules coordinated to OH$^-$ (H$_2$O⋯OH$^-$) and Na$^+$ (H$_2$O⋯Na$^+$). The water orientation curve shows a slightly negative orientation, in agreement with literature[44]. The few OH$^-$ anions within 3.5 Å from the instantaneous water interface show a clear net orientation with their hydrogen atoms pointing towards the vapor phase. However, its contribution to the overall signal is minor, because only very few OH$^-$ groups are present at this depth. Indeed, most OH$^-$ anions are over 3.5 Å below the instantaneous water interface, without a net orientation. The orientation of the coordinated water is consistent with the predicted VSFG spectra: H$_2$O⋯OH$^-$ molecules point, on average, down to the solution, while H$_2$O⋯Na$^+$ water molecules are oriented upwards, towards the interface. Importantly, most of the net orientation of the coordinated builds up within a region void of Na$^+$ and OH$^-$ ions.

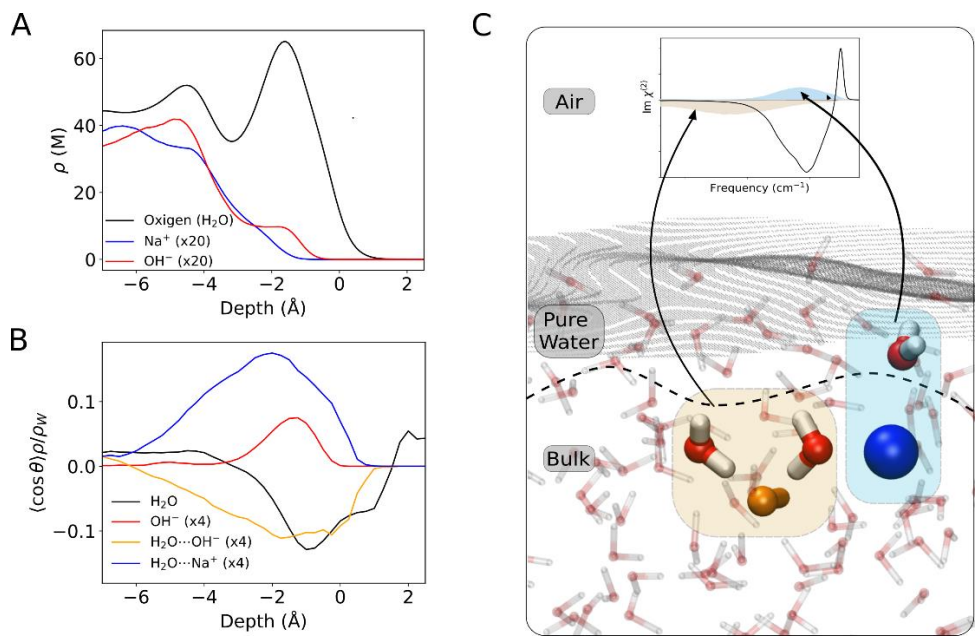

**Figure 3: Microscopic analysis of the air/NaOH(aq) interface**. Depth profiles of (A) the density profiles and (B) the average angle of the H-O-H bisector of water molecules and OH⁻ for the 1.25M NaOH solution. The origin point is defined as the position of the instantaneous liquid interface. (**A**) Oxygen atoms of the water molecules, Na+, and OH- are plotted in black, blue, and red colors, respectively. (**B**) Water molecules without ions in its vicinity (black), OH⁻ (red), water molecules in the vicinity of Na⁺ ions (blue), and water molecules in the vicinity of OH⁻ ions (orange). A negative (positive) <cosθ> indicates a net *down*-orientation towards the bulk of the solution (*up*-orientation towards the air). Values are scaled by its density relative to the water density. Some curves are further scaled for clarity. (**C**) A representative snapshot of the MD simulations. Na⁺ and OH⁻ are highlighted in blue and orange, respectively. Selected water molecules coordinated to the ions which are responsible for the observed changes in the HD-VSFG spectrum are also highlighted. The instantaneous air/liquid interface is depicted by a gray surface and the dashed lines represent the interfaces between pure water and bulk electrolyte solution. The H₂O···OH⁻ and H₂O···Na⁺ water molecules giving rise to the changes in the VSFG spectra are marked with light-blue and orange areas, respectively.

With all the previous observations in mind, we propose the following two-interfaces model depicted in Figure 3c (MD snapshot). The lack of ions on the topmost layer creates a thin internal layer of 'pure' water which gives rise to effectively two interfaces: an air/water interface and a water/electrolyte solution interface. The water molecules inside this thin layer which are next to ions, show the specific orientation along the surface normal accordingly, causing the observed changes in the $\text{Im}(\chi^{(2)})$ spectra. On the one hand, in the case of $Na^+$ ions, due to the spherical symmetry of the cation, only one orientation is favored in which the oxygen atoms reside closest to the cation. On the other hand, and as mentioned above, most $OH^-$ ions are located 3.5 Å below the instantaneous air/water interface. The hydrogen atoms of the $OH^-$ motif does not act as hydrogen-bond donor[40], while the oxygen atom serves as a strong hydrogen bond acceptor being able to accept between 3 and 5 hydrogen bonds per atom[38]. Thus, on average, $OH^-$ ions align more water molecules pointing toward the solution than to the interface. Last but not least, inside the bulk region, the orientations get randomized, producing a vanishing net orientation and therefore no additional SFG signal. In section 4 in the SI, we present a simple one-dimensional model that showcases how the absence of ions in one region of space can lead to a distinct water alignment, as we propose here.

Until this point, we have restricted our focus to the aqueous NaOH solution. If the existence of the liquid/liquid interface was a general phenomenon for other electrolytes, it would be possible to evidence it from spectroscopic measures as well. Thus, to verify our new microscopic interpretation, we carried out HD-VSFG measurements for a series of halogen salts with varying strength of the water-anion interaction.

We show the HD-VSFG spectra for HCl, NaCl, NaBr, and NaI aqueous interfaces in Figure 4a. Since the spectral changes induced by the addition of ions are proportional to their concentration (see Figures S1-S6 in the SI), the total signal can be decomposed into a linear combination of the water, $\chi_w^{(2)}$, and $\chi_{electrolyte}^{(2)}$, as shown in Eq. 1. Thus, Figure 4b should, in principle, originate from a superposition of a positive signal emerging from the water aligned towards the interface due to its interaction with Na$^+$ ions ($\chi_c^{(2)}$), with a negative signal arising from the water alignment induced by the anions ($\chi_a^{(2)}$). For NaOH, these different contributions are nicely separated, owing to the greater frequency separation between the different water molecules near cations and halide anions.

To test this assumption, we approximate the $\chi_c^{(2)}$ spectral shape by the constructed $(I_{IR} * I_{Raman})^{1/2}$ spectrum, where $I_{IR}$ and $I_{Raman}$ represent the IR and Raman intensities, respectively[45]. Since the water O−H stretch spectrum is governed by the intermolecular interaction acting on the hydrogen atom of water[46], and the Na$^+$ ion barely affects the O−H stretch mode, the Na$^+$ induced perturbation should, in principle, resemble the spectral features present in bulk water. This assumption is validated by the resemblance between the constructed $(I_{IR} * I_{Raman})^{1/2}$ spectrum and the spectra shown in Figure 1c (see also Figure S7).

The obtained $\chi_a^{(2)}$ spectra are displayed in Figure 4c. The negative contribution follows the expected trend and gets red-shifted along the series I$^-$ → Br$^-$ → Cl$^-$ → F$^-$, which follows the decrease of the relative strength of the corresponding hydrogen bonds[47]. In accordance with the proposed modeling, stronger anion-water interactions give rise to a relatively stronger negative contribution. Similar to what is observed in gas phase clusters[48,49], the F$^-$ anion forms a remarkably strong hydrogen bond which blue shifts the O−H stretch spectrum by more than 200 cm$^{-1}$ compared to the other halides.

We stress that the EDL interpretation, which attributes the alignment to the local field generated by the ion separations and disregards explicit water-ion interactions, cannot explain the changes observed in the differential spectra for these electrolytes (Figure 4b).

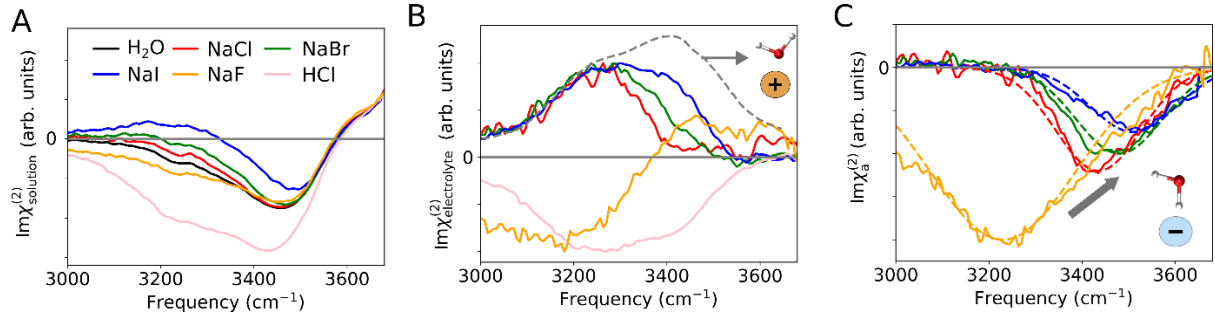

Figure 4: **Deconvolution of HD-VSFG into Cationic- and Anionic-induced Spectral Contributions (A)** Imaginary part of HD-VSFG spectra, $\text{Im}(\chi^{(2)}_{\text{solution}})$, obtained for H$_2$O (black), NaF 1M (orange), NaCl 1M (red), NaBr 1M (green), NaI 1M (blue), and HCl 1M (pink) aqueous solutions. **(B)** Electrolyte contribution to the second order response computed as $\chi^{(2)}_{\text{electrolyte}} = \chi^{(2)}_{\text{solution}} - \chi^{(2)}_{\text{water}}$. The signals are normalized to the maximum value. The geometric mean of Raman and IR signals for H$_2$O, $(I_{IR} * I_{Raman})^{1/2}$, is depicted by grey dashed line. **(C)** Estimated anionic contribution computed as $\chi^{(2)}_{a} = \chi^{(2)}_{\text{electrolyte}} - b(I_{IR} * I_{Raman})^{1/2}$, where $b$ is a scaling factor. A Gaussian fit to guide the eye are shown as dashed line.

In Figure 4, we also show data from the HCl solution. As mentioned above, its $\chi^{(2)}_{\text{electrolyte}}$ spectrum shows a monotonic change characterized by a decrease of the signal in the H-bonded O−H stretch region below 3600 cm$^{-1}$. Together with the fact that the free O−H signal decreases markedly with concentration, see Figure 1e, we believe that this is the only system, among the electrolytes considered here, for which the EDL model needs no rectification and provides the correct microscopic picture. While we suspect that the water coordinate to the Cl$^-$ anion should principle create a negative contribution, as it does for the other cases, the strong signal arising from proton-induced alignment masks it entirely. In Figure S8 in the SI, we also report the data obtained for CsF. The determined $\chi^{(2)}_{\text{electrolyte}}$ shows positive and negative regions, which resemble the NaOH and NaF cases. While the EDL model fails again to rationalize simultaneous spectral changes with different signs, the liquid-liquid interface picture gives a clear interpretation: the positive and negative peaks arise from water coordinated to the cations and anions, respectively. The deconvolution of the $\chi^{(2)}_a$ for the F$^-$ anion agrees with the one obtained for NaF (Figure 4c) and demonstrates the robustness of the proposed analysis. Finally, to verify the emerging picture, we further performed *ab initio* MD simulations for the sodium halide salts and computed the corresponding density profiles. As shown in Figure S11, the air/liquid interface of the NaF, NaCl and NaBr solutions is depleted of ions leading to the aforementioned interfacial stratification. This effect is less pronounced for the NaI solution, which appears as an intermediate situation between the rest of the halides and the proton.

In summary, HD-VSFG, in combination with *ab initio* simulations, in this case, aided by NN, has proven to be an invaluable tool to contribute to the molecular-level understanding of liquid surfaces. Unlike other techniques, such as electron spectroscopy[50] and second harmonic generation[51], which provide a relatively coarse-grained view, the vibrational resolution of heterodyne-detected VSFG enables an important revision of the established EDL picture. The presence or absence of ions in the outermost surface layer determines their availability to participate in chemical reactions as well as their reactivity[52,53]. The emerging picture of stratification of electrolyte solutions at the interface substantially expands on current textbook descriptions, providing powerful insights toward resolving the air/water interface puzzle.

## Methods

**Ab initio simulations.** Density functional theory (DFT) calculations were carried out with the CP2K package[54] at the revPBE-D3(0) level of theory. This choice has been shown to provide an excellent compromise in terms of the description quality of interfacial water and the required computational cost[55]. We employed the TZV2P basis sets. The core electrons were described by the Goedecker-Teter-Hutter pseudopotential, and the real-space density cut-off was set to 320 Ry. We used a simulation cell with dimensions 16.63 x 16.63 x 44.10 Å. NaOH 1.25M, 2.50M, and 3.75M solutions were modeled by a simulation cell containing a slab made of 160, 156, and 152 water molecules, and 4, 8 and 12 NaOH molecules, respectively. The slab thickness was found to be approximately 18±1 Å in all the cases. The simulations for NaF, NaCl, NaBr, and NaI were performed in larger simulation cells with dimensions 14.4 x 14.4 x 70.0 Å and were modelled by a slab made of 248 water molecules and 6 sodium halide molecules. In these cases, the slab thickness was found to be approximately 38±1 Å. The molecular dynamics simulations for the sodium halide salts were performed with the CP2K package. For each salt, we prepared 10 different initial coordinates and carried out a total of 400 ps in the NVT ensemble controlled by a stochastic velocity rescaling thermostat ($\tau$ = 300 fs) with a target temperature of 300K to ensure the correct equilibration of the ion distribution.

**Neural network training.** The training of the Behler-Parinello high-dimensional neural network (NN) was performed using the n2p2 code [56,57] following the active-learning strategy outlined in Ref. [58]. The reference data were obtained from two independent 20 ps long NVE trajectories for each concentration, previously thermalized at 300K. We started the training with 40 randomly picked structures and trained six different NN models to form a committee model[59]. The disagreement among the NN models was used to identify the most relevant configurations, which were added by batches of 20, till the disagreement across could not be improved further. In total, we collected around 300 structures where we obtained energy and force root mean square errors for both training and test sets below 1 meV per atom and 100 meV/Bohr, respectively. These overall accuracy estimates are known to yield an accurate representation of the reference potential energy surface[32]. It is noteworthy to mention that the NN architecture employed in this work does not include long-range effects explicitly. However, since we used a rather small slab, and the 6 Å cut-off employed to describe the atomic environments, which effectively becomes 12 Å in the force calulations[32], most part of the electrostatic interactions can be described adequately[60]. Indeed, similar architecture has been used to model bulk aqueous NaOH solutions and obtained very good agreement with experimental results[61–63].

**Neural network molecular dynamics simulations.** The production runs for the NaOH solutions were performed with the i-pi program[64] connected to the LAMMPS package[57,65]. For each concentration, we prepared 10 different initial coordinates and carried out a total of 1 ns in the NVT ensemble controlled by a stochastic velocity rescaling thermostat[66] ($\tau$ =300 fs) with a target temperature of 300K to ensure the correct equilibration of the ion distribution. The instantaneous interfaces were calculated using the Willard-Chandler method[67]. The surface-specific velocity-velocity correlation function[30] (ssVVCF) were calculated using a Hahn window of 1 ps and obtained by averaging 10 independent 100 ps trajectories carried out from uncorrelated initial coordinates extracted from the NVT trajectories. The time step employed for the integration of the equation of motions was in all cases 0.5 fs. Unless stated otherwise, the classification as either OH⁻ or water molecule was performed by assuming that each H is covalently bound exclusively to its nearest oxygen atom[61]. Finally, we considered a water molecule to be coordinated to OH⁻ ($H_2O\cdots OH^-$) or to Na⁺ ($H_2O\cdots Na^+$) if the oxygen-oxygen distance is below 3.0 Å or if the oxygen-sodium distance is below 3.5 Å, respectively. We verified that slightly different cut-off distances did not qualitatively affect the reported results.

**Sample preparation.** HCl (36.5%) was obtained from Alfa Aesar. NaOH (>98%), NaF (>99%), NaCl (>99%), NaBr (>99%), NaI (>99%), CsF (>99%), and CsCl (>99%,) were obtained from Sigma Aldrich. These chemicals except NaF, NaCl, and NaBr were used without further purification in the study. NaF, NaCl, and NaBr were baked in an oven for 8 hours at 650 °C before experiments to remove organic contaminants. Pure water was obtained from a Milli-Q system (the resistance of 18.2MΩ cm). By dissolving the salt with pure water, we obtained various concentrations of salt solutions. NaI solution was freshly prepared 1 hour before measurement to avoid oxidation of iodide ion. Before measurement, all solutions were covered with aluminum foil to prevent the photochemical reaction from light. All the measurements are performed under nitrogen purge conditions.

**Heterodyne-detected sum frequency generation spectroscopy.** For NaOH aqueous solution samples, we use a non-collinear beam geometry with a Ti:Sapphire regenerative amplifier laser system (Spitfire Ace, Spectra-Physics, centered at 800 nm, ~40 fs pulse duration, 5 mJ pulse energy, 1 kHz repetition rate). A part of the output was directed to a grating-cylindrical lens pulse shaper to produce a narrowband visible pulse (8 µJ pulse energy at the sample position, FWHM = ~10 cm$^{-1}$), while the other part was used to generate a broadband infrared (IR) pulse (3 µJ pulse energy at the sample position, FWHM = ~400 cm$^{-1}$) through an optical parametric amplifier (OPA, Light Conversion TOPAS-C) with a silver gallium disulfide (AgGaS$_2$) crystal. The IR and visible beams were firstly focused onto a 200 nm-thick ZnO on a 1 mm-thick CaF$_2$ window to generate a local oscillator (LO) signal in a similar manner to Ref. [68]. Three beams were collimated with an off-axis parabolic mirror. A fused silica glass plate with a 1.5 mm thickness was placed in the optical path for the LO signal, allowing the phase modulation for the LO signal. Then, visible, LO, and IR beams were re-focused by an off-axis parabolic mirror at incident angles of 64°, 61°, and 50° at the sample solution interface, respectively. The SFG signal from the sample interfered with the SFG signal from the LO, generating the SFG interferogram. The SFG interferogram was then dispersed into a spectrometer (Shamrock 303i, Andor Technology) and detected by a CCD camera (Newton, Andor Technology). To monitor the height change of the sample surface due to water evaporation, we used a height displacement sensor (CL-3000, Keyence).

For other electrolyte solutions, i.e., NaCl, NaBr, NaI, CsF, and HCl samples, we measured the HD-VSFG spectra on a collinear beam geometry using a Ti: Sapphire regenerative amplifier (Spitfire Ace, Spectra-Physics, centered at 800 nm, ~40 fs pulse duration, 5 mJ pulse energy, 1 kHz repetition rate). A part of the output was used to generate a broadband IR pulse in an OPA with an $AgGaS_2$ crystal. The other part of the output was directed through a pulse shaper consisting of a grating-cylindrical mirror system to generate a narrow band visible pulse with a bandwidth of ~13.5 $cm^{-1}$. The IR and visible beam were firstly focused into a 20 μm-thick y-cut quartz plate as the LO. Then these beams were collinearly passed through a 5 mm thick $SrTiO_3$ plate for the phase modulation and focused into on the sample surface at angles of incidence of 45°. The SFG signal from the sample interfered with the LO signal, generating the SFG interferogram. The SFG interferogram was dispersed into a spectrometer (Teledyne Princeton Instruments, HRS-300) and detected by a liquid-nitrogen cooled CCD camera (Teledyne Princeton Instruments, PyLoN).

The complex spectra of second-order nonlinear susceptibility ($\chi^{(2)}$) were obtained via the Fourier analysis of the SFG interferogram and normalized by a z-cut quartz crystal. Unless stated otherwise, all the measurements were performed with *ssp* (denoting *s-*, *s-*, and *p-* polarized SFG, visible, and IR beams, respectively) polarization combination. For our analysis, we used $(\chi^{(2)})_{ssp}$ data without the Fresnel factor correction.

## Data Availability

The data corresponding to Figures 1, 2, 3a, 3b, 4, S1-S9 and S11, input files to reproduce the *ab initio* and NN simulations will become available at
https://gitlab.com/litman90/electrolytesolutions-si
at the moment of publication.

## Code Availability

All the code used in this work is freely available and can be found in the corresponding references provided in Methods.

68. Vanselous, H. & Petersen, P. B. Extending the Capabilities of Heterodyne-Detected Sum-Frequency Generation Spectroscopy: Probing Any Interface in Any Polarization Combination. *The Journal of Physical Chemistry C* **120**, 8175–8184 (2016).
## Acknowledgments

We thank Maksim Grechko and Stephen J. Cox for insightful discussions and helpful comments on the manuscript and the MaxWater Initiative of the Max Planck Society for financial support.
## Author Contributions

Y.L., K-Y. C., T. S., Y. N., and M. B. designed the research. Y.L. performed the simulations and K-Y. C. and T. S. performed the experimental measurements and analyzed the data. All authors discussed the results and Y.L. wrote the initial version of the manuscript. All authors revised, contributed, and proofread the final manuscript and the Supplementary Information.

## Competing interests

The authors declare no competing interests.

# Supplementary Material: The surface of electrolyte solutions is stratified

Yair Litman,[1,2*] Kuo-Yang Chiang,[1] Takakazu Seki,[1]

Yuki Nagata,[1] Mischa Bonn[1,*]

Max Planck Institute for Polymer Research,
Ackermannweg 10, 55128 Mainz, Germany

Yusuf Hamied Department of Chemistry, University of Cambridge,
Lensfield Road, Cambridge, CB21EW, United Kingdom

*To whom correspondence should be addressed;
E-mail: litmany@mpip-mainz.mpg.de, bonn@mpip-mainz.mpg.de

# 1 Salt-concentration-dependent sum-frequency generation spectra of electrolyte solutions

To obtain the complex $\chi^{(2)}$ spectra for NaCl, NaBr, NaI, CsF, and HCl in Fig. 4 of the main text, we performed heterodyne-detected sum-frequency generation (HD-VSFG) measurement at the air/electrolyte interfaces by varying the concentrations of ions. The Im$\chi^{(2)}$ data are present in panel (a) of Figs. S1-S5. We observed a negative Im$\chi^{(2)}$ band between 3000 cm$^{-1}$ and 3600 cm$^{-1}$ for the pure water case. The addition of NaCl, NaBr, and NaI elevate the Im$\chi^{(2)}$ signal below 3400cm$^{-1}$, while the addition of HCl and CsF enhance the negative contribution in the <3400cm$^{-1}$ region.

We further calculate the electrolyte spectra defined as, $\chi^{(2)}_{\text{electrolyte}} = \chi^{(2)}_{\text{solution}} - \chi^{(2)}_{\text{water}}$, to examine whether the ion-induced spectral changes are constant throughout all concentration regions. Panel (b) of Figs. S1-6 shows the $\text{Im}\chi^{(2)}_{\text{electrolyte}}$ signals for various ions. The amplitude of $\chi^{(2)}_{\text{electrolyte}}$ increases with increasing salt concentrations. Nevertheless, the spectral shapes remain unchanged, as shown by the normalized spectra presented in panel c of Figs. S1-6. This manifests that the structure of the interfacial water changes independently of the ion concentrations considered here.

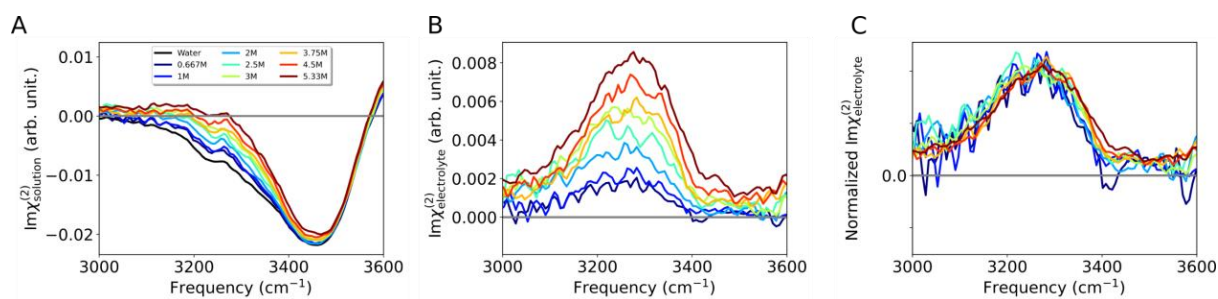

Figure S1. **HD-VSFG spectra for various concentrations of NaCl solution**. (**A**) Imaginary part of $\chi^{(2)}$ spectra, $\text{Im}\chi^{(2)}_{\text{solution}}$ (**B**) The difference spectra, $\text{Im}\chi^{(2)}_{\text{electrolyte}}$ (**C**) Normalized $\text{Im}\chi^{(2)}_{\text{electrolyte}}$ to the same scale for comparison.

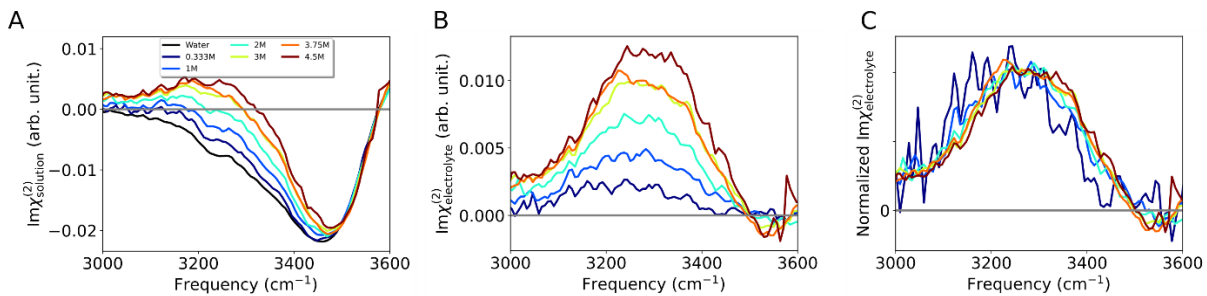

Figure S2. **HD-VSFG spectra for various concentrations of NaBr solution**. (**A**) Imaginary part of $\chi^{(2)}$ spectra, Im $\chi^{(2)}_{\text{solution}}$ (**B**) The difference spectra, Im $\chi^{(2)}_{\text{electrolyte}}$ (**C**) Normalized Im$\chi^{(2)}_{\text{electrolyte}}$ to the same scale for comparison.

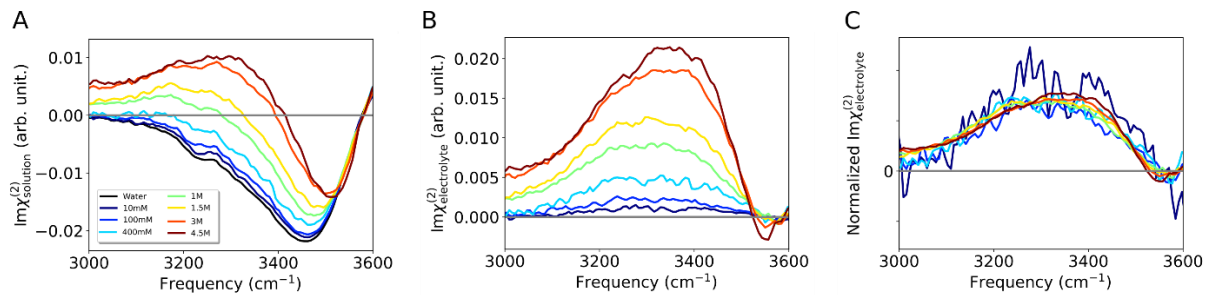

Figure S3. **HD-VSFG spectra for various concentrations of NaI solution.** (**A**) Imaginary part of $\chi^{(2)}$ spectra, Im $\chi^{(2)}_{\text{solution}}$ (**B**) The difference spectra, Im $\chi^{(2)}_{\text{electrolyte}}$ (**C**) Normalized Im$\chi^{(2)}_{\text{electrolyte}}$ to the same scale for comparison.

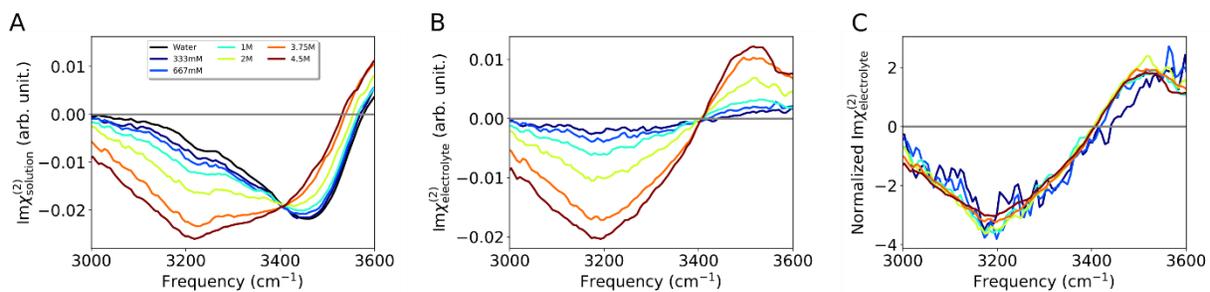

Figure S4. **HD-VSFG spectra for various concentrations of CsF solution** (**A**) Imaginary part of $\chi^{(2)}$ spectra, Im $\chi^{(2)}_{\text{solution}}$ (**B**) The difference spectra, Im $\chi^{(2)}_{\text{electrolyte}}$ (**C**) Normalized Im$\chi^{(2)}_{\text{electrolyte}}$ to the same scale for comparison.

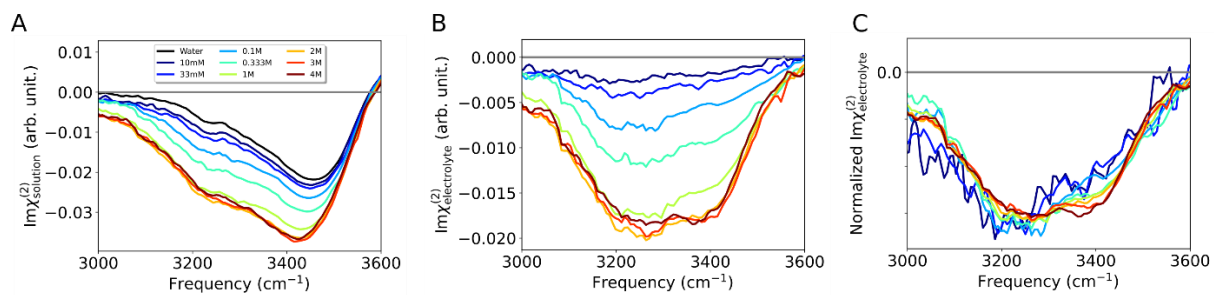

Figure S5. **HD-VSFG spectra for various concentrations of HCl solution** (**A**) Imaginary part of $\chi^{(2)}$ spectra, Im $\chi^{(2)}_{\text{solution}}$ (**B**) The difference spectra, Im $\chi^{(2)}_{\text{electrolyte}}$ (**C**) Normalized Im$\chi^{(2)}_{\text{electrolyte}}$ to the same scale for comparison.

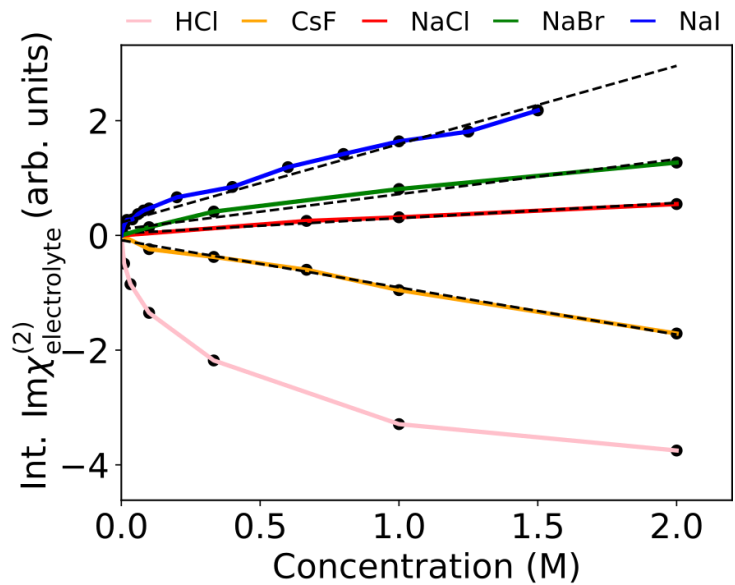

Figure S6: **Concentration dependence of Im $\chi^{(2)}_{electrolyte}$**. Concentration dependence of integrated Im$\chi^{(2)}_{electrolyte}$ signals for HCl (pink), CsF (orange), NaCl (red), NaBr (green), and NaI (blue). Linear fits are depicted by dashed black lines. In all the cases the signal is integrated from 3000 to 3350 cm$^{-1}$.

## 2  Further analysis on ion contributions to Im$\chi^{(2)}$

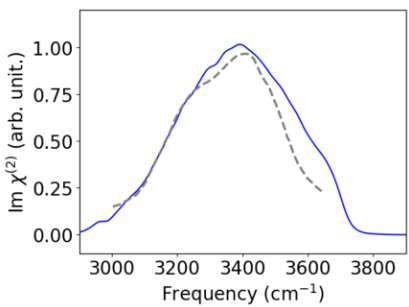

Figure S7. **Comparison of Experimental $(I_{IR} * I_{Raman})^{1/2}$ with Simulated Na$^+$ induced Spectral Contribution**. Contribution to the simulated Im$\chi^{(2)}_c$ spectra by the water molecules in the vicinity of Na$^+$ ions for NaOH (blue) and geometric mean of geometric mean of Raman and IR signals of pure water (gray dashed). The former is displaced by -50 cm$^{-1}$.

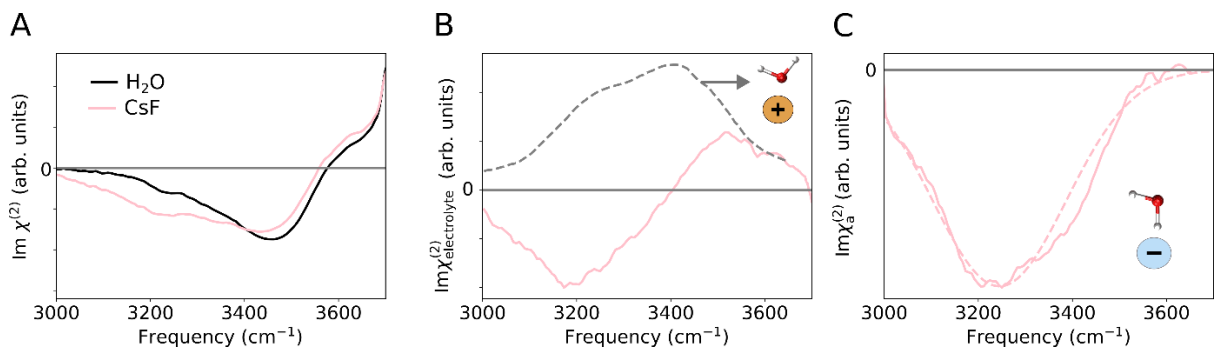

Figure S8: **Deconvolution of HD-VSFG into Cs$^+$ and F$^-$ induced Spectral Contributions (A)** Imaginary part of VSFG spectra obtained for H$_2$O (black) and NaF (pink) aqueous solutions. **(B)** Electrolyte contribution computed as $\chi^{(2)}_{\text{electrolyte}} = \chi^{(2)}_{\text{solution}} - \chi^{(2)}_{\text{water}}$. The geometric mean of Raman and IR signals for H$_2$O, $(I_{IR} * I_{Raman})^{1/2}$, is depicted by grey dashed line. **(C)** Estimated anionic contribution computed as $\chi^{(2)}_{\text{a}} = \chi^{(2)}_{\text{electrolyte}} - b(I_{IR} * I_{Raman})^{1/2}$, where $b$ is a scaling factor. A Gaussian fit to guide the eye are shown as dashed line.

# 3 NN validation tests

Besides the global error obtained on the predicted energies and forces reported in the Methods section, we verified that quality of the predicted velocity density of states (VDOS). In Fig. S9, we compare the predicted VDOS by the NN and the reference data. Even though the reference curves present a relatively large noise due to the small simulation length, a very good agreement between the NN and *ab initio* simulations is obtained.

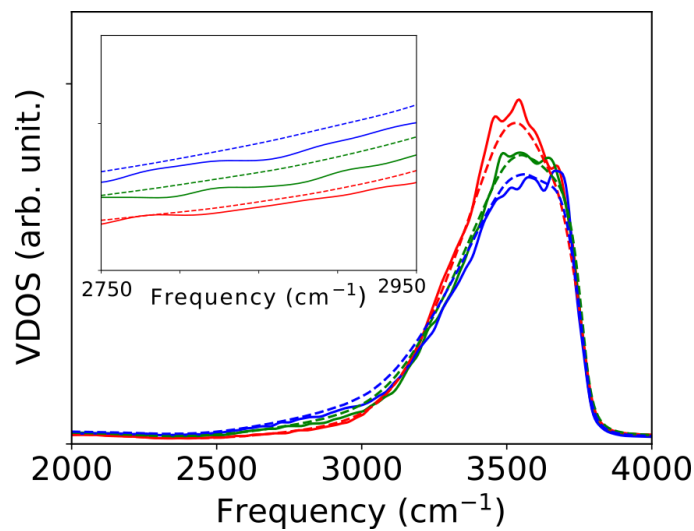

Figure S9. **Vibrational density of states (VDOS) of NaOH(aq) slab structures.** VDOS of NaOH aqueous solutions obtained with the reference *ab initio* trajectories (solid lines) and the NN (dashed lines). NaOH 1.25 M, NaOH 2.50 M, and NaOH 3.75 M are depicted with red, green, and blue colors, respectively.

# 4 Illustration of the Interface Stratification by a Simple 1D Model

In this section we describe a simple 1D model to represent the geometrical effects caused by the stratification of the water interface. We consider a 1D slab of length **L** with an inner bulk-like region of length **B**. The bulk region contains water molecules, cations and anions. Importantly, while the ions are constrained between **-B/2** and **B/2**, the water molecules can be found outside of this region. We assume that each ion is solvated by two water molecules at each side and that water-ion distance is **d**. The numerical simulation goes as follows:

1) Pick a random position between **-B/2** and **B/2** to place the ion and name it **X**
2) Add +1/-1 at a distance **X**+/-**d**. This represents the orientation and position of the water molecules (Note the water molecules can fall outside the [**-B/2,B/2**] region)

3) Repeat steps 1 and 2 until convergence of the corresponding histogram.

In Fig. S10a and Fig. S10b, we present two cartoons describing one step of numerical modelling for an anion and a cation, respectively. In Figure S10c, we show the results after 20000 iterations. First, the counts in the **[-D/2+d,+D/2+d]** regions averages to zero implying that the lack of net orientation of water molecules inside the bulk-like region. Second, looking at the position **D/2**, positive and negative net orientation are built for the water coordinating cation and anion, respectively. Due to symmetry, the opposite effect is observed at the position **-D/2**. However, in macroscopic samples, where **D** is several micrometres, experiments can easily discriminate both interfaces. Finally, since the vibrational frequencies of water molecules coordinated to anions and cations are normally different, the positive and negative signals emerging from the different 'types' of water molecules do not cancel out and generate a characteristic signature in a real spectrum.

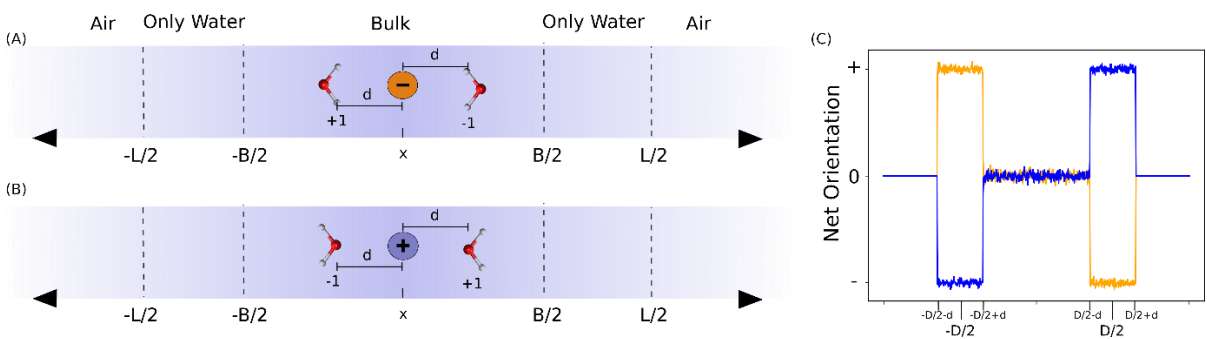

Figure S10: **Description of the 1D model illustration of the stratification at the water/air interface. (A)** Cartoon representing one step of the numerical modelling for an anion **(B)** Same as a for a cation **(C)** Result of the numerical model after 20000 iterations. Anions and cations curves are represented by blue and orange lines, respectively.

# 5 Density Profiles of Sodium Halides Salts

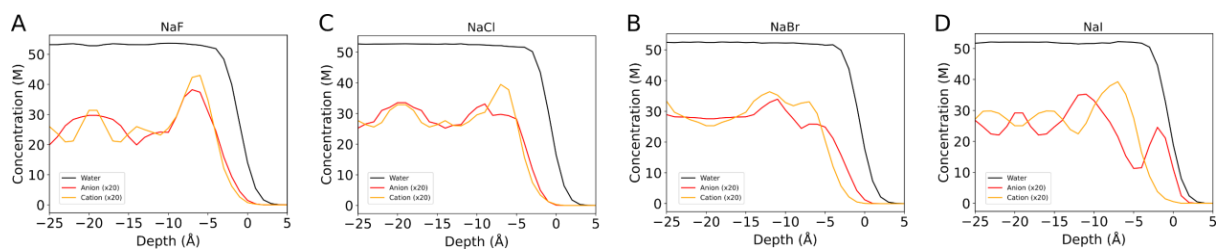

Figure S11: **Average density profiles.** Average density profiles of **(A)** NaF, **(B)** NaCl, **(C)** NaBr, and **(D)** NaI obtained from *ab initio* MD simulations.